\documentclass[fleqn,10pt]{wlscirep}
\usepackage[utf8]{inputenc}
\usepackage[T1]{fontenc}
\title{Decoherence by Optical Phonons in GaN Defect Single-Photon Emitters}

\author[1,*]{Yifei Geng}
\author[2]{Jialun Luo}
\author[3]{Len van Deurzen}
\author[1,4]{Huili (Grace) Xing}
\author[1,4]{Debdeep Jena}
\author[3]{Gregory David Fuchs}
\author[1]{Farhan Rana}

\affil[1]{School of Electrical and Computer Engineering, Cornell University, Ithaca, New York 14853, USA.}
\affil[2]{Department of Physics, Cornell University, Ithaca, New York 14853, USA.}
\affil[3]{School of Applied and Engineering Physics, Cornell University, Ithaca, New York 14853, USA.}
\affil[4]{Department of Materials Science and Engineering, Cornell University, Ithaca, New York 14853, USA.}

\affil[*]{yg474@cornell.edu}

\begin{abstract}
  Single-photon defect emitters (SPEs), especially those with magnetically and optically addressable spin states, in technologically mature wide bandgap semiconductors are attractive for realizing integrated platforms for quantum applications. Broadening of the zero phonon line (ZPL) caused by decoherence in solid state SPEs limits the indistinguishability of the emitted photons. Decoherence also limits the use of defect states in quantum information processing, sensing, and metrology. In most defect emitters, such as those in SiC and diamond, interaction with low-energy acoustic phonons determines the temperature dependence of the decoherence rate and the resulting broadening of the ZPL with the temperature obeys a power law. GaN hosts bright and stable single-photon emitters in the 600 nm to 700 nm wavelength range with strong ZPLs even at room temperature. In this work, we study the temperature dependence of the ZPL spectra of GaN SPEs integrated with solid immersion lenses with the goal of understanding the relevant decoherence mechanisms. At temperatures below  $\sim$50 K, the ZPL lineshape is found to be Gaussian and the ZPL linewidth is temperature independent and dominated by spectral diffusion. Above $\sim$50 K, the linewidth increases monotonically with the temperature and the lineshape evolves into a Lorentzian. Quite remarkably, the temperature dependence of the linewidth does not follow a power law. We propose a model in which decoherence caused by absorption/emission of optical phonons in an elastic Raman process determines the temperature dependence of the lineshape and the linewidth. Our model explains the temperature dependence of the ZPL linewidth and lineshape in the entire 10 K to 270 K temperature range explored in this work. The $\sim$19 meV optical phonon energy extracted by fitting the model to the data matches remarkably well the $\sim$18 meV zone center energy of the lowest optical phonon band ($E_{2}(low)$) in GaN. Our work sheds light on the mechanisms responsible for linewidth broadening in GaN SPEs. Since a low energy optical phonon band ($E_{2}(low)$) is a feature of most group III-V nitrides with a wurtzite crystal structure, including hBN and AlN, we expect our proposed mechanism to  play an important role in defect emitters in these materials as well.  
\end{abstract}
\begin{document}

\flushbottom
\maketitle

\thispagestyle{empty}

\section*{Introduction}

Single-photon emitters (SPEs) play an important role in quantum computing and communication technologies~\cite{aharonovich2016solid}. On-demand solid-state single-photon emitters have been demonstrated in different material systems including semiconductor quantum dots~\cite{claudon2010highly,santori2001triggered}, color centers in wide bandgap materials such as diamond~\cite{kurtsiefer2000stable,neu2012photophysics} and SiC~\cite{castelletto2014silicon}, and defects in two-dimensional materials~\cite{he2015single,tran2016quantum}. The identification of high-brightness, spectrally pure, and high-efficiency SPEs in technologically mature semiconductor systems that are compatible with high-quality epitaxial growth, offer suitable refractive index contrasts for photonic devices, and enable integration with control electronics is highly desirable~\cite{aharonovich2016solid}.  Recently, defect-based SPEs in AlN~\cite{xue2020single} and GaN~\cite{berhane2017bright,berhane2018photophysics} have been reported. GaN is a direct, wide bandgap material of high technological significance in applications related to visible wavelength lasers and light emitting diodes, and semiconductor RF and power devices. SPEs in GaN are therefore interesting and technologically relevant. GaN SPEs were reported to be bright, photostable, and exhibited sharp photoluminescence (PL) peaks spread out in the 600-700 nm wavelength range~\cite{berhane2017bright,berhane2018photophysics}. The nature of these GaN SPEs remains elusive. Point defects in GaN as well as electron states localized at stacking faults and dislocations in the crystal have been proposed as candidates~\cite{Nguyen2019point,Nguyen2021polarity}.

In this work, we study the temperature dependence of the ZPL emission spectra in GaN SPEs and propose a novel decoherence mechanism involving interaction with optical phonons to be responsible for the observed ZPL linewidth broadening. ZPL linewidth broadening caused by decoherence is a challenge for the generation of indistinguishable photons needed in many quantum systems. The temperature dependence of the zero phonon line (ZPL) emission spectrum provides a wealth of information not just about the nature of defect-based SPEs but also offers a window into the physical processes responsible for decoherence and emission linewidth broadening. In most solid state defect SPEs, interaction with low energy acoustic phonons is responsible for the temperature dependence of the decoherence rates as well as emission linewidth broadening. Various physical models for the decoherence induced by acoustic phonon have been proposed to explain the temperature dependence of the emission linewidths observed in solid state SPEs. For example, the $T^{3}$ temperature dependence observed in AlN, SiC and hBN SPEs~\cite{xue2020single,sontheimer2017photodynamics,lienhard2016bright} has been attributed to acoustic phonon-induced dephasing in crystals with a large number of defects~\cite{Reineker1999T3}. The $T^{5}$ dependence observed in NV$^{-}$ centers in diamond has been shown to result from the dynamic Jahn-Teller effect in the excited state~\cite{Fu2009JT,abtew2011dynamic}. The $T^{7}$ dependence observed in many solid state emitters has been attributed to quadratic coupling to acoustic phonons~\cite{hizhnyakov2002zero,Silsbee1962T7}. Interaction with optical phonons is generally not considered an important mechanism for decoherence at temperatures much below room temperature given the large energies of the optical phonons.  

Our experimental results show that at temperatures below $\sim$50 K, the ZPL has a Gaussian lineshape and the linewidth saturates at values in the 0.7-1 meV range (0.2-0.3 nm range). This low temperature linewidth is attributed to spectral diffusion. As the temperature increases, the ZPL lineshape evolves from a Gaussian into a Lorentzian. Interestingly, the temperature dependence of the linewidth does not follow any of the power laws that work for many other solid state SPEs (discussed above). We propose a model in which decoherence and linewidth broadening occurs by absorption/emission of optical phonons in an elastic Raman process. The linewidth data matches the model closely and the optical phonon energy extracted by fitting the model to the data is found to be $\sim$19 meV, a value that matches the energy of the lowest $E_{2}(low)$ Raman-active optical phonon band in GaN remarkably well. Our work helps to elucidate the nature of the SPEs in GaN and the physics associated with their decoherence as a result of defect-phonon interactions. Since a low energy optical phonon band ($E_{2}(low)$) is a feature of most group III-V nitrides with a wurtzite crystal structure, including AlN and hBN, we expect our proposed decoherence mechanism to play an important role in defect emitters in these materials as well. In fact, recent works on hBN SPEs have already pointed out the absence of power law temperature dependence of the ZPL linewidth~\cite{jungwirth2016temperature,akbari2021temperature}.

\section*{Results}

\subsection*{GaN defect single photon emitters}

\begin{figure}[tbp]
        \centering
		\includegraphics[width=0.6\linewidth]{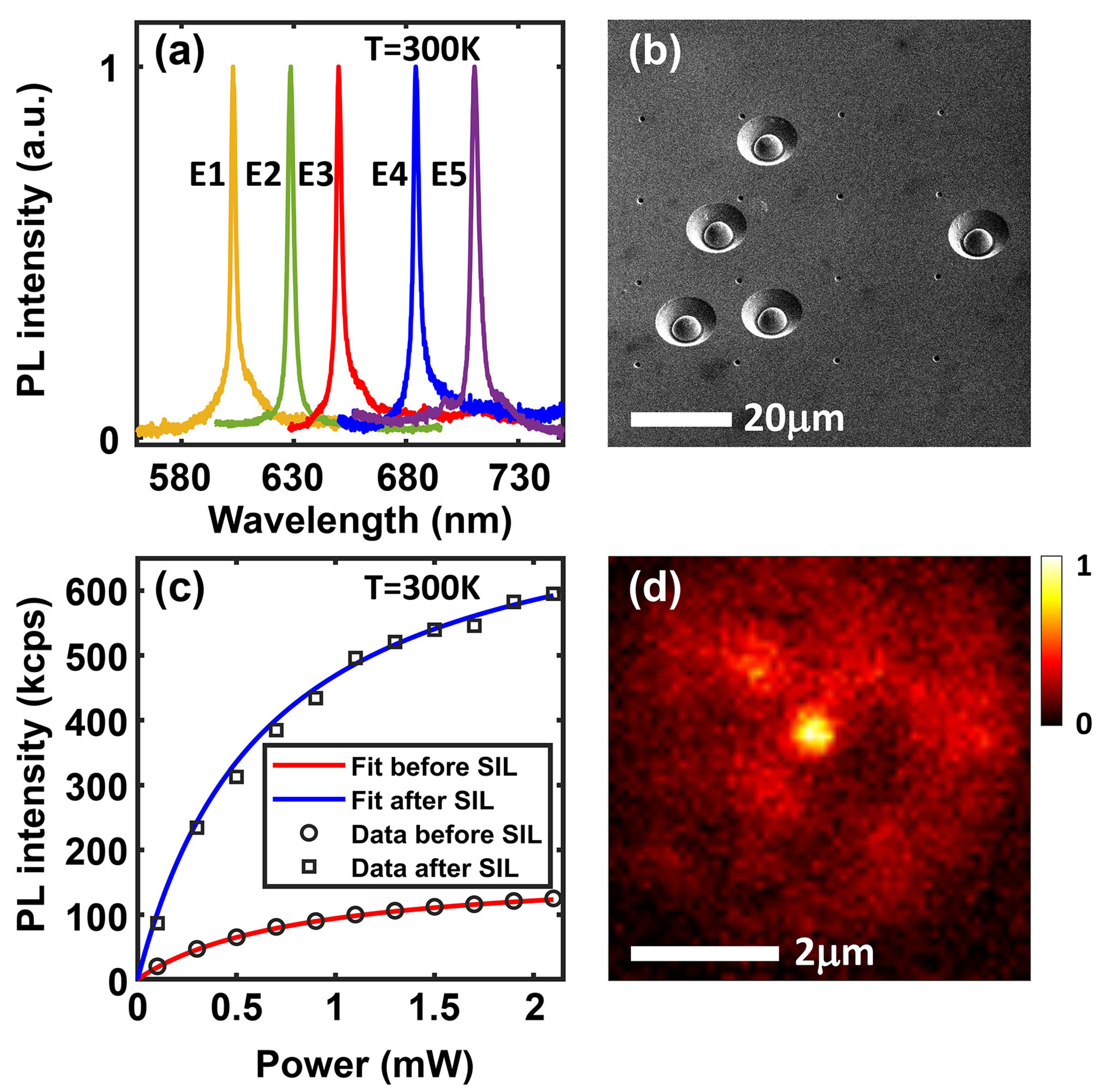}
		\caption{ (a) Representative PL spectra of five GaN SPEs, E1 through E5, are plotted at room temperature. (b) SEM image of five solid immersion lenses (SILs), each fabricated around a SPE, is shown. Each SIL is a hemisphere of radius 2.5 $\mu$m. (c) The measured PL intensities of a SPE before and after the fabrication of a SIL are plotted as a function of the pump power. (d) The spatial PL map of a single emitter within a SIL is shown.}
		\label{fig:sil}
	\end{figure}

In this work, we investigate single-photon defect emitters in HVPE grown GaN epitaxial layers. GaN defect emitters exhibit strong ZPLs in the 600-700 nm wavelength range at room temperature. Representative emission spectra of some SPEs are shown in Fig.~\ref{fig:sil}(a). The ZPL center wavelengths of emitters E1 through E5 are 602.9 nm, 628.7 nm, 650.1 nm, 684.5 nm and 710.5 nm, respectively. These wavelengths match well with the wavelengths for GaN SPEs reported earlier~\cite{berhane2017bright}.

GaN is a high index material in the visible wavelength range. As a consequence, most PL is trapped inside the substrate due to total internal reflection. To increase the photon collection efficiency, a solid immersion lens (SIL)~\cite{marseglia2011nanofabricated,jamali2014microscopic} in the form of a hemisphere of radius 2.5 $\mu$m was fabricated on top of each emitter by focused ion beam milling of GaN, as shown in Fig.~\ref{fig:sil}(b). To avoid deflection of the ion beam due to surface charge accumulation during milling, a 30 nm Al layer was sputtered on the GaN surface before milling and Al left after milling was removed using a wet etch. PL collection efficiency from the SPEs was found to be enhanced by factors in the 4-5 range (using a 0.9 NA objective). Fig.~\ref{fig:sil}(c) shows the PL intensity (in kcps) of an emitter before and after the fabrication of a SIL as a function of the pump power. Fig.~\ref{fig:sil}(d) shows the PL map of a defect emitter in the center of the SIL. The measured PL intensity $I_{pl}$ can be fitted by the standard relation, 
\begin{equation}
    I_{pl} = I_{sat}\frac{P_{pump}}{P_{pump}+P_{sat}}
\end{equation}
Here, $I_{sat}$ is the saturation PL intensity, $P_{pump}$ is the pump power, and $P_{sat}$ is the saturation pump power. For the data shown in Fig.~\ref{fig:sil}(c), $P_{sat}$ is 650 $\mu$W, $I_{sat}$ is 171 kcps without the SIL and 779 kcps with the SIL, indicating that the PL collection efficiency is enhanced by a factor of $\sim$4.5. This enhancement in light collection ensured a sufficient signal-to-noise ratio for cryogenic temperature measurements when a smaller NA (0.7) objective was used.

\begin{figure}[tbp]
        \centering
		\includegraphics[width=0.6\linewidth]{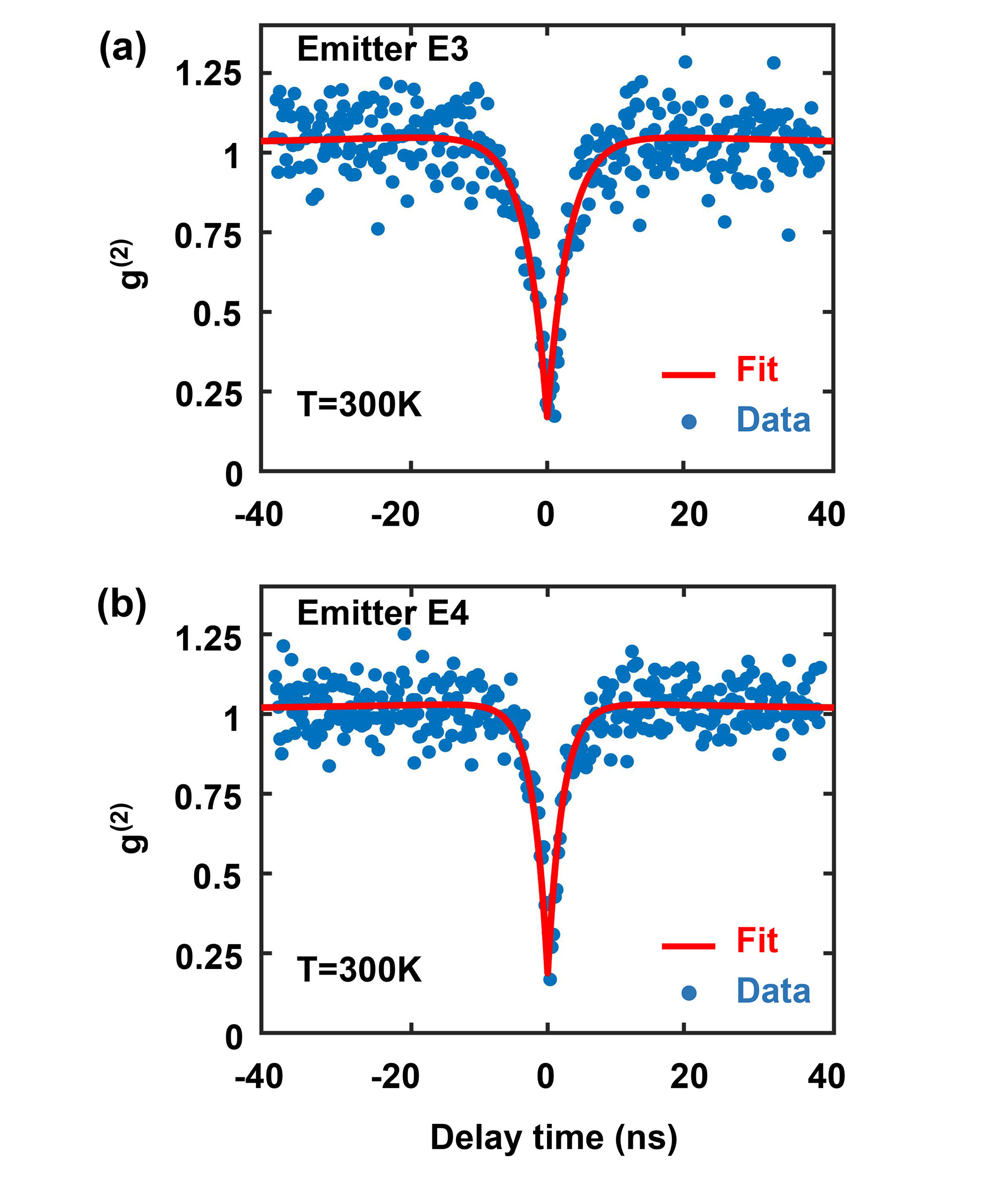}
		\caption{(a) The second order correlation function $g^{(2)}(\tau)$ of emitter E3 is plotted. $g^{(2)}(0)=0.17$. (b) $g^{(2)}(\tau)$ of emitter E4 is plotted. $g^{(2)}(0)=0.19$. The solid lines show the fits obtained using the expression given in the text.}
		\label{fig:g2}
\end{figure}

In what follows, we focus mostly on two emitters, E3 and E4 in Fig.~\ref{fig:sil}(a), with center emission wavelengths 650.1 nm ($\sim$1907.4 meV) and 684.5 nm ($\sim$1811.5 meV), respectively. Most other emitters were found to exhibit characteristics similar to them. Fig.~\ref{fig:g2} shows the measured second order correlation function $g^{(2)}(\tau)$ for these two emitters at room temperature using a pump power of 50 $\mu$W. $g^{(2)}(\tau)$ was obtained using the time-tagged time-resolved (TTTR) mode of the MultiHarp150 instrument and was properly normalized. For both emitters, $g^{(2)}(\tau)$ can be fitted by the expression,
\begin{equation}
    g^{(2)}(\tau)=1-ae^{-|\tau |/\tau_{1}}+be^{-|\tau |/\tau_{2}}
\end{equation}
The fits are shown by the solid lines in Fig.~\ref{fig:g2}. The extracted values of $\tau_{1}$ are $3.18\pm 0.24$ ns and $2.2\pm 0.17$ ns for emitters E3 and E4, respectively, and the values of $\tau_{2}$ are $74\pm 27$ ns and $65\pm 33$ ns for emitters E3 and E4, respectively. $g^{(2)}(0)$ equals 0.17 and 0.19 for emitters E3 and E4, respectively, which confirms these defects as single-photon emitters. The measured values of $\tau_{1}$ and $\tau_{2}$ are in good agreement with the values reported previously~\cite{berhane2017bright}.

\begin{figure}[tbp]
        \centering
		\includegraphics[width=0.6\linewidth]{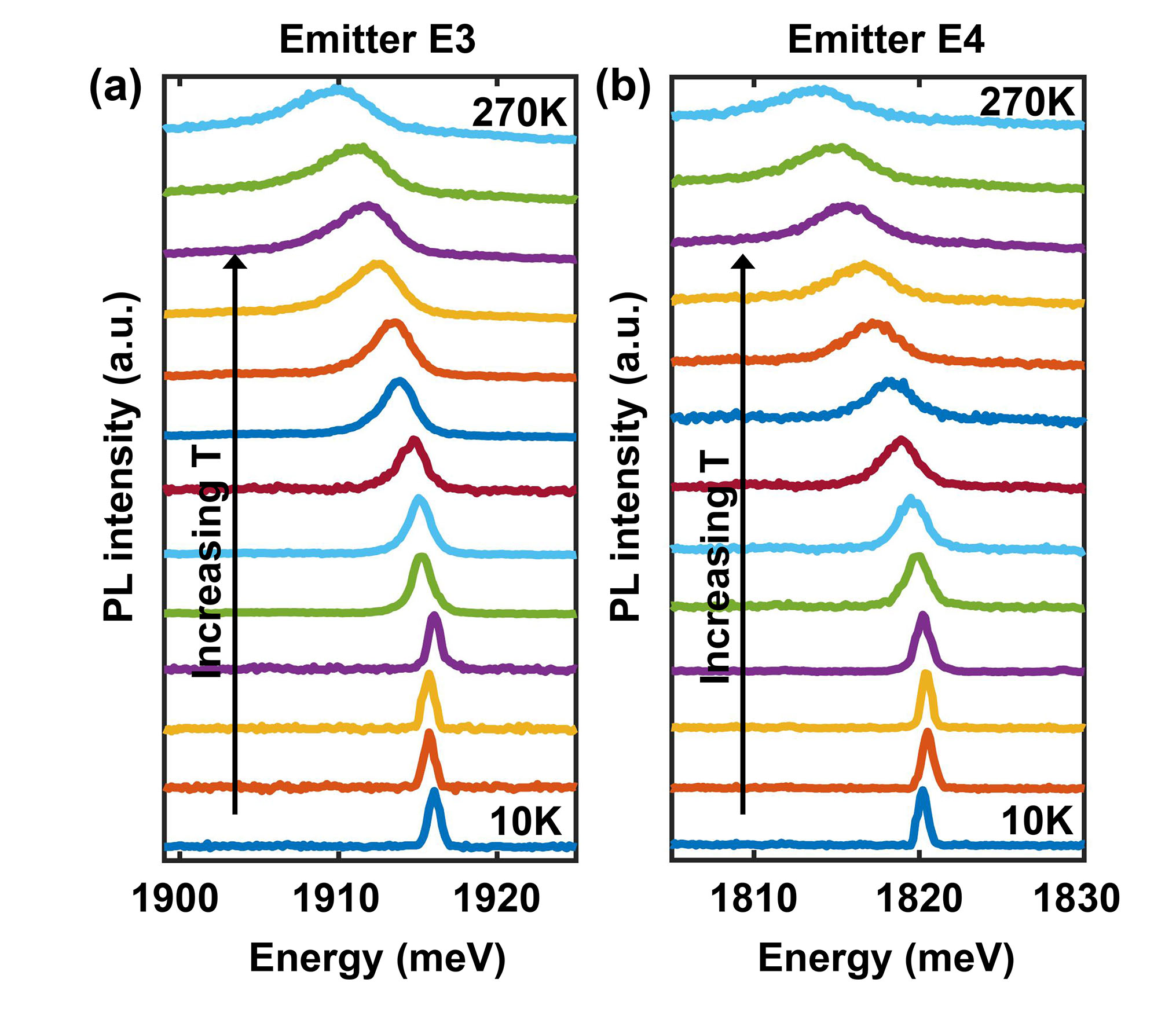}
		\caption{The emission spectra of emitter E3 (a) and E4 (b) are plotted for different temperatures in the 10 K to 270 K range (with an increment of 20 K).}
		\label{fig:allT spectra}
\end{figure}

\subsection*{Temperature dependence of the ZPL emission spectra}

The ZPL emission spectra were measured for temperatures in the 10 K to 270 K range and the results are shown in Fig.~\ref{fig:allT spectra} for emitters E3 and E4. Other emitters display similar trends. The center emission energies are seen to redshift with an increase in the temperature. The ZPL energy of E3 shifts from 1916 meV at 10 K to 1909.4 meV at 270 K. In the case of E4, the ZPL energy shifts from 1820.2 meV at 10 K to 1813.5 meV at 270 K. We don't observe a S-shaped temperature dependence of the ZPL center energies reported previously~\cite{berhane2017bright}.

\begin{figure}[tbp]
        \centering
		\includegraphics[width=0.6\linewidth]{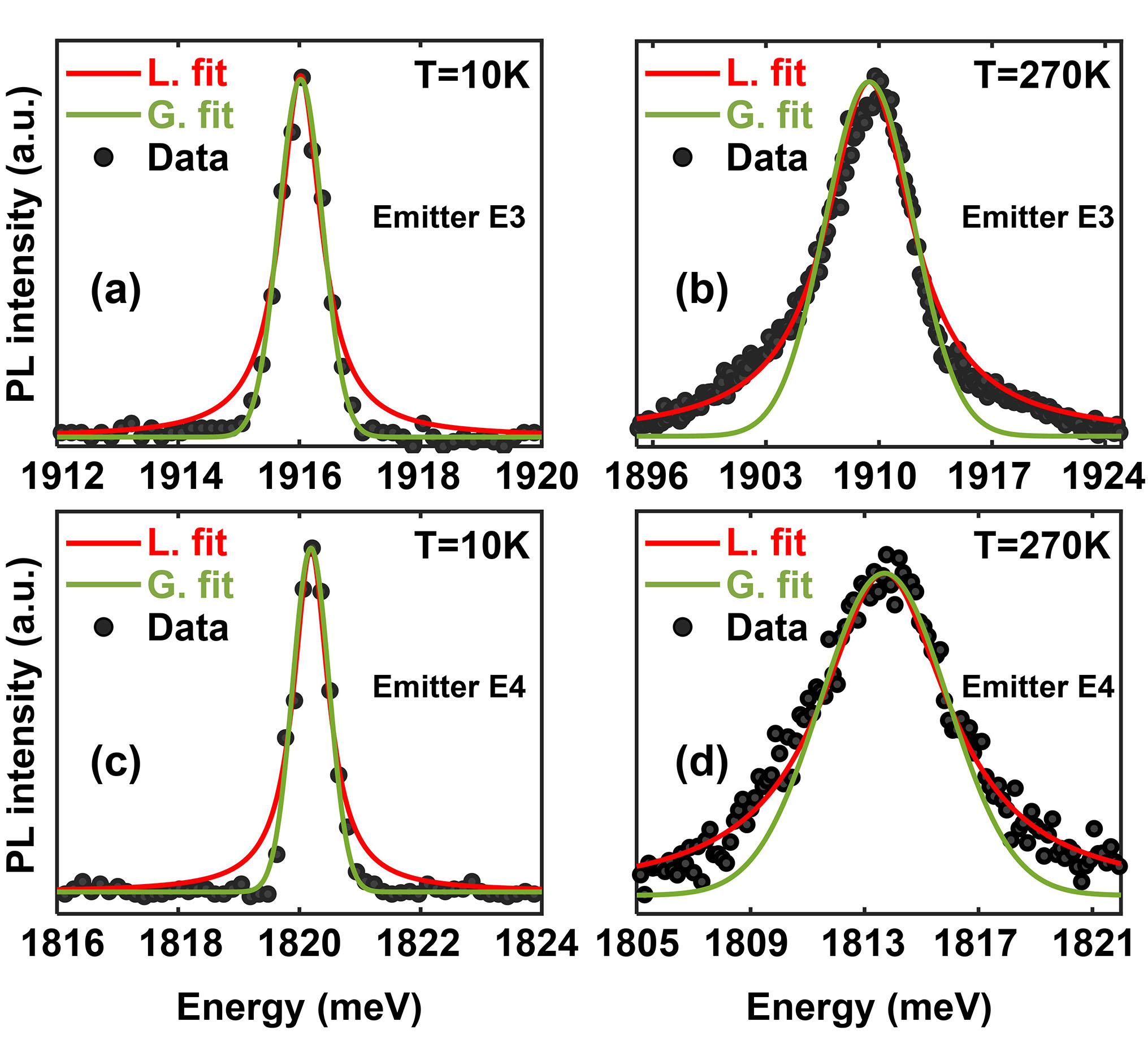}
		\caption{The ZPL spectra with a Gaussian and a Lorentzian fits at 10 K (a) and 270 K (b) for emitter E3 are plotted. Also shown are the ZPL spectra with a Gaussian and Lorentzian fits at 10 K (c) and 270 K (d) for emitter E4.}
		\label{fig:spectra of E3 and E4}
\end{figure}

\begin{figure}[tbp]
        \centering
		\includegraphics[width=0.6\linewidth]{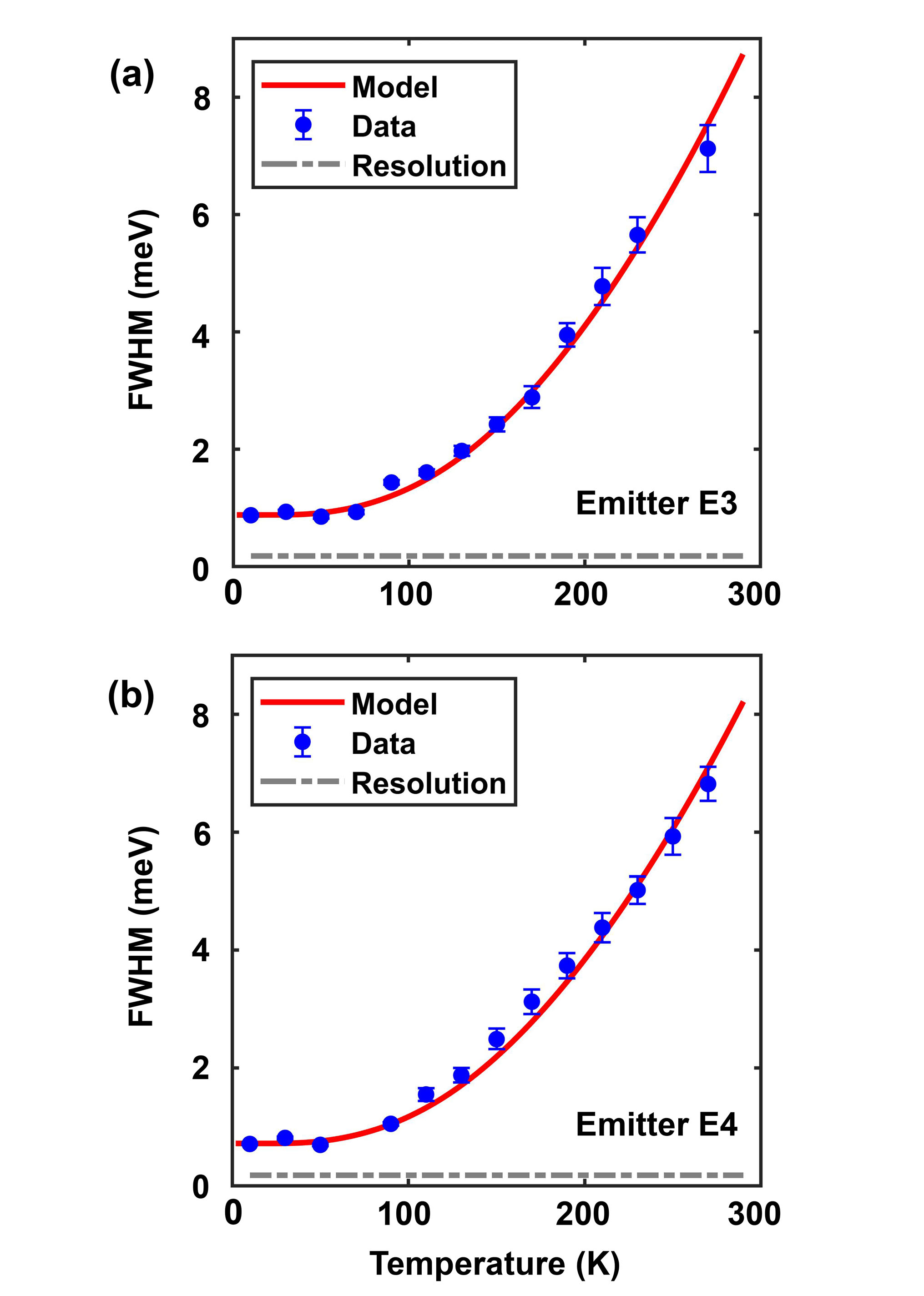}
		\caption{The FWHM linewidths of emitter E3 (a) and emitter E4 (b) are plotted as a function of the temperature. The solid lines are the fit to the data using the theoretical model discussed in the text.}
		\label{fig:tdependence}
\end{figure}

We look at the spectral shape of the ZPL as a function of the temperature. Our data shows that the ZPL spectra evolve from a Gaussian lineshape at temperatures below $\sim$50 K to a Lorentzian lineshape at temperatures above $\sim$125 K. This is shown in detail in Fig.~\ref{fig:spectra of E3 and E4} which plots the ZPL spectra of emitters E3 and E4 at low (10 K) and high (270 K) temperatures, along with Gaussian and Lorentzian fits to these spectra at the two temperatures. The data shown was obtained using a pump power of 300 $\mu$W. At 10 K, the spectrum of E3 (E4) can be fit much better with a Gaussian spectral function with a full-width-half-maximum (FWHM) linewidth of 0.88 meV (0.72 meV). Whereas at 270 K, the spectrum of E3 (E4) can be fit much better with a Lorentzian spectral function with a FWHM linewidth of 7.12 meV (6.82 meV). These observations suggest that two different mechanisms are contributing to the linewidth. One may make the simplest assumption that these two mechanisms are independent. Under this assumption, the ZPL spectral shape is more accurately given by a Voigt function $V(\omega;\sigma,\gamma)$ which is a convolution of Gaussian and Lorentzian functions~\cite{sontheimer2017photodynamics,akbari2021temperature,neu2013low},       
\begin{equation}
    V(\omega;\sigma,\gamma) \propto \int_{-\infty}^{+\infty}G(\omega^{'};\sigma)L(\omega-\omega^{'};\gamma)d\omega^{'}
\end{equation}
Here, $G(\omega;\sigma)$ and $L(\omega;\gamma)$ are Gaussian and Lorentzian functions with FWHM equal to $f_{G} = 2\sigma\sqrt{2\ln2}$ and $f_{L} = 2\gamma$, respectively. The FWHM $f_{V}$ of the Voigt function can be written as,
\begin{equation}
    f_V=0.5346f_L+\sqrt{0.2166 f_{L}^{2}+f_{G}^{2}} \label{eq:fv}
\end{equation}
By fitting the measured ZPL spectra with a Voigt function, the temperature dependent FWHM of its Gaussian and Lorentzian components can be extracted. We find that the FWHM $f_{G}$ of the Gaussian component is temperature independent and is around 0.88 meV (0.72 meV) for E3 (E4). An Emission spectrum with a temperature independent FWHM and a Gaussian lineshape is a common signature of spectral diffusion whereby the emitter emission energy changes in time as a result of factors such as changes in the electrical environment of the emitter. To gain insight into the mechanism responsible for the Lorentzian component, which dominates at temperatures higher than $\sim$125 K, we look at the FWHM linewidth of the ZPL as a function of the temperature. This data is shown in Fig.~\ref{fig:tdependence}(a) for emitter E3 and in Fig.~\ref{fig:tdependence}(b) for emitter E4. It is clear that the Lorentzian component dominates the ZPL linewidth at temperatures higher than $\sim$125 K. Using the expression given in Eq.(\ref{eq:fv}), we find that the temperature dependence of the FWHM of the Lorentzian component cannot be adequately fitted with an expression proportional to $T^{n}$, where $n$ is any integer greater than or equal to 3 ($n$ equals 3, 5, and 7 for some common decoherence mechanisms mentioned earlier in this paper). Fig.\ref{fig:app} in the Appendix shows the poor comparison with the data that is obtained if the temperature dependence of the linewidth of the Lorentzian component is assumed to be $T^{3}$. This indicates that the common decoherence mechanisms, discussed earlier in this paper, might not be the dominant decoherence mechanisms in the case of GaN SPEs. Given that the measured ZPL linewidths for GaN SPEs, shown in Fig.~\ref{fig:tdependence}, are orders of magnitude larger than the linewidths that would result from relaxation processes (as estimated by the measured $g^{(2)}$ function), the decoherence mechanism is quite strong. Below, we present a theoretical model for decoherence and show that this model fits the data very well.

\subsection*{A theoretical model for decoherence in G\MakeLowercase{a}N single-photon emitters}

\begin{figure}[tb]
        \centering
		\includegraphics[width=0.8\linewidth]{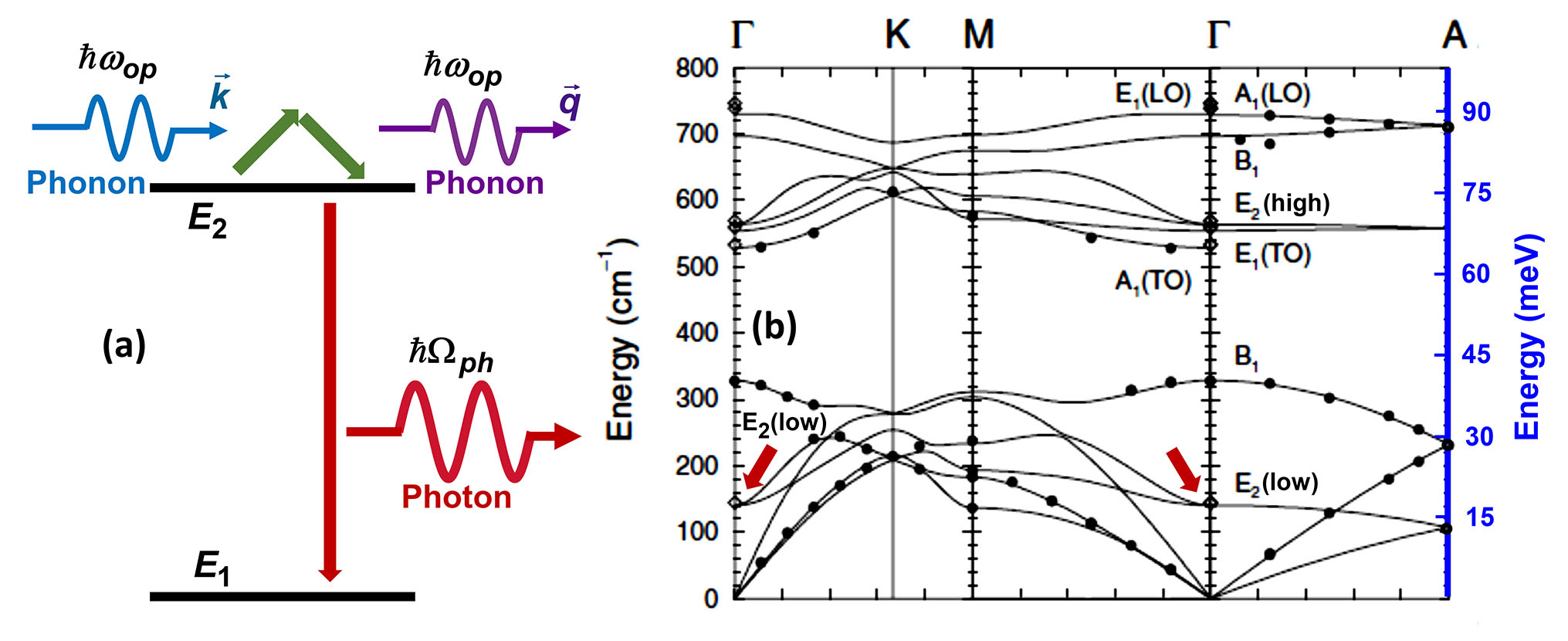}
		\caption{(a) Proposed mechanism for decoherence involving absorption/emission of optical phonons via an elastic Raman process. $E_{2}$ ($E_{1}$) represents the excited (ground) state energy of the emitter. (b) Phonon bands in wurtzite GaN are reproduced from the work of Ruf et al.~\cite{Cardona2001phonon}. The energy of the $E_{2}(low)$ optical phonon mode at the zone center matches the value obtained by fitting the model to the data.}
		\label{fig:raman}
\end{figure}

The decoherence mechanism proposed here is depicted in Fig.~\ref{fig:raman}(a) and it involves absorption/emission of optical phonons in an elastic Raman process which results in the scattering of an optical phonon from the defect. A similar mechanism involving acoustic phonons is known to result in a decoherence rate proportional to $T^{7}$ in solid state emitters~\cite{Silsbee1962T7,hizhnyakov2002zero}. Although as depicted in Fig.~\ref{fig:raman}(a) the decoherence occurs in only the excited state, decoherence occurring by a similar process in the ground state can be handled in a way similar to the one shown below.

We assume that the Hamiltonian for the defect state interacting with optical phonons is,
\begin{eqnarray}
  & & H = \sum_{j} \left[ E_{j} + \alpha_{j} F(t) \right] c_{j}^{\dagger}c_{j} \nonumber \\
  & & + \frac{1}{\sqrt{V}} \sum_{j \ne 2,\vec{k}}  (M_{j,\vec{k}}c_{j}^{\dagger}c_{2} + M^{*}_{j,-\vec{k}}c_{2}^{\dagger}c_{j})(a_{\vec{k}} + a_{-\vec{k}}^{\dagger}) \nonumber \\
  & & + \frac{1}{\sqrt{V'}} \sum_{\vec{k}} (F_{\vec{k}}c_{1}^{\dagger}c_{2} + F^{*}_{-\vec{k}}c_{2}^{\dagger}c_{1})(b_{\vec{k}} + b_{-\vec{k}}^{\dagger}) \nonumber \\
  & & + \sum_{\vec{k}} \hbar \omega_{\vec{k}} a_{\vec{k}}^{\dagger}a_{\vec{k}} + \sum_{\vec{k}} \hbar \Omega_{\vec{k}} b_{\vec{k}}^{\dagger}b_{\vec{k}} 
\end{eqnarray}
Here, $c_{j}$, $a_{\vec{k}}$, and $b_{\vec{k}}$ are the destruction operators for the electron, optical phonon, and photon states, respectively. $E_{j}$ are the energies of the emitter electron states and, as shown in Fig.~\ref{fig:raman}, $j=1,2$ states participate in photon emission. $\omega_{\vec{k}}$ are the frequencies of the optical phonons in a band that is coupled to the emitter. $\Omega_{\vec{k}}$ are the frequencies of the photon modes. $V$ ($V'$) is the volume in which the phonon (photon) modes are normalized. To model spectral diffusion, we have included terms in the electron energies linear in the external time-dependent electric field $F(t)$ which is assumed to be caused by time-dependent charges in the environment. Terms quadratic in $F(t)$ can also be included in  the Hamiltonian but their inclusion does not affect the discussion that follows and the conclusions~\cite{Santis2021stark}. We assume that $\langle F(t) \rangle = 0$ and $\langle F(t) F(t')\rangle = F_{o}^{2} e^{-\lambda |t-t'|}$, where $\lambda^{-1}$ is the field correlation time and will be assumed to be much longer than any other time scale in the problem. The electron-phonon interaction term in the Hamiltonian couples the emitter excited state to virtual states that can be adiabatically eliminated to give the following effective electron-phonon interaction Hamiltonian for the process shown in Fig.~\ref{fig:raman}(a),
\begin{equation}
  H^{eff}_{e-ph} = c_{2}^{\dagger}c_{2} \frac{1}{V} \sum_{\vec{k} \ne \vec{q}} G_{\vec{k},\vec{q}} a_{\vec{q}}^{\dagger} a_{\vec{k}}
\end{equation}
where,
\begin{equation}
G_{\vec{k},\vec{q}} =  \sum_{j \ne 2} \left[ \frac{M^{*}_{j,\vec{q}}M_{j,\vec{k}}}{E_{2}-E_{j} + \hbar \omega_{\vec{k}}} + \frac{M^{*}_{j,-\vec{k}}M_{j,-\vec{q}}}{E_{2}-E_{j} - \hbar \omega_{\vec{q}}} \right]  
\end{equation}
If the electron-phonon interaction is via optical deformation potential then it is reasonable to assume that $G_{\vec{k},\vec{q}}$ will be large only when both $\vec{k},\vec{q}$ are small (near the center of the Brillouin zone)~\cite{Ziman}. The ZPL emission spectrum $S(\omega)$ can be obtained from the relation~\cite{Milburn},
\begin{eqnarray}
  S(\omega) & = & \int dt \, e^{-i\omega t} \langle \sigma_{+}(t) \sigma_{-}(0) \rangle / \langle c_{2}^{\dagger}c_{2} \rangle \nonumber \\
  & = & \int dt \, e^{-i\omega t} \langle c_{2}^{\dagger}(t) c_{1}(t) c_{1}^{\dagger}c_{2} \rangle /\langle c_{2}^{\dagger}c_{2} \rangle
\end{eqnarray}
Using the cumulant expansion technique for the quantum propagator, the above expression gives,
\begin{equation}
  S(\omega) \approx \int dt \, e^{-i\left[\omega - (E_{2}-E_{1})/\hbar\right] t} \, e^{-\sigma^{2}t^{2}/2} e^{-(\gamma + \gamma_{sp}) |t|} \label{eq:S}
\end{equation}
Here, $\sigma = |\alpha_{2}-\alpha_{1}|F_{o}/\hbar$, $2\gamma_{sp}$ is the spontaneous emission rate, and the decoherence rate $\gamma$ due to interaction with phonons is,
\begin{eqnarray}
  \gamma & = & \frac{2\pi}{\hbar^{2}} \int d\omega D^{2}(\omega) |G(\omega)|^{2} \, n(\omega)\left[n(\omega) + 1 \right] \nonumber \\
  & \approx & \frac{2\pi}{\hbar^{2}} n(\omega_{op})\left[n(\omega_{op}) + 1 \right] \int d\omega D^{2}(\omega) |G(\omega)|^{2} \label{eq:gamma}
\end{eqnarray}
$n(\omega)$ is the thermal boson occupation factor and $D(\omega)$ is the density of states function for the optical phonons. In the writing the result in Eq.(\ref{eq:S}), we have ignored the shift in the energy $E_{2}$ that results from phonon and photon interactions. We will assume that $\gamma >> \gamma_{sp}$ and that the decoherence is almost entirely due to interaction with phonons. The product $D^{2}(\omega) |G(\omega)|^{2}$ inside the integral is assumed to be peaked near the frequency $\omega_{op}$. Eq.(\ref{eq:S}) shows that the ZPL spectral shape will be given by a Voigt function. The expression for $\gamma$ shows that the temperature dependence of the decoherence rate is determined by the product $n(\omega_{op})\left[n(\omega_{op}) + 1 \right]$, which gives a temperature dependence very different from any power law. 

Using the expression for the FWHM of $S(\omega)$ given earlier in Eq.(\ref{eq:fv}) with the experimentally determined values of the Gaussian component $f_{G}$, and using the temperature dependence of $\gamma$ given by the expression in Eq.(\ref{eq:gamma}) for the Lorentzian component $f_{L}=2\gamma$, we can fit the measured FWHM of the ZPL, for both emitters E3 and E4, over the entire 10 K to 270 K temperature range with a root mean square error less than 0.05 meV provided we assume that $\hbar \omega_{op}$ equals 19 meV $\pm$0.5 meV. The fits obtained for $\hbar \omega_{op} = 19$ meV are shown by the solid lines in Fig.~\ref{fig:tdependence}. The excellent agreement between the data and the model begs the question if 19 meV is close to any one of the bulk optical phonons energies in GaN. Quite remarkably, the lowest energy Raman-active $E_{2}(low)$ optical phonon band in GaN has energy equal to $\sim$18 meV at the $\Gamma$-point of the Brillouin zone, as shown in Fig.~\ref{fig:raman}(b)~\cite{Cardona2001phonon}. Since, as stated earlier, $|G(\omega)|^{2}$ is expected to be large near the zone center, the experimental value of 19 meV for $\hbar \omega_{op}$ is reasonable and consistent with decoherence being caused by the coupling between the emitter and the bulk $E_{2}(low)$ optical phonons. The $E_{2}(low)$ optical phonon in GaN is known to be Raman active and couples strongly to the electronic states in the valence and conduction bands~\cite{Phillips2011}.

\section*{Discussion}

Since decoherence rate due to the process in Fig.\ref{fig:raman}(a) is proportional to $n(\omega_{op})\left[n(\omega_{op}) + 1 \right]$, the rate would have have been negligibly small, especially at low temperatures, if it were not for the fact that $\hbar \omega_{op}$ is also very small. Nitrides, and GaN in particular, are quite unique among wide bandgap semiconductors in that they possess optical phonon modes with low energies at the Brillouin zone center and these optical phonons modes are Raman-active and couple to the electronic states. Even if the emitter is coupled to other higher energy optical phonons, one would expect the lowest energy optical phonon with the largest thermal occupation to contribute the most to the decoherence rate via the mechanism shown in Fig.~\ref{fig:raman}(a) and this is also consistent with our data.

The coupling of the GaN defect SPEs to the low energy bulk optical phonon band is interesting because it suggests that the crystal lattice structure is not distorted by the defects to the extent that the bulk phonon modes become significantly modified in the vicinity of the defects. Furthermore, the presence of a sharp and strong ZPL at even room temperature, in contrast to the ZPLs of many other defects (e.g. NV$^{-}$ centers in diamond) that are visible at only low temperatures, suggests that a localized optical phonon mode at the defect site is either absent or is very weakly coupled to the emitter (i.e. a small Huang-Rhys factor). Finally, the thermal stability of the defect SPEs suggests that these defects are very likely not interstitials. The above characteristics are all consistent with the SPEs being substitutional impurity atoms or substitutional impurity-vacancy complexes. We should mention here that recently electron states localized at stacking faults and dislocations in the crystal have also been proposed as candidates for these SPEs~\cite{Nguyen2021polarity}. Clearly, more work is needed to determine the nature of GaN SPEs.

\section*{Conclusion}

In conclusion, we have investigated GaN SPEs and studied the temperature dependence of their emission spectrum. In contrast to previous reports, we find that both the FWHM ZPL linewidth as well as the emission center wavelength increase monotonically with the temperature. The temperature dependence of the ZPL linewidth can be explained very well over the entire 10 K to 270 K temperature range by our proposed model in which decoherence occurs via absorption/emission of optical phonons. The experimentally determined optical phonon energy matches well the zone center energy of the lowest optical phonon band ($E_{2}(low)$) in GaN. Bright, stable, and fast, GaN SPEs have the potential for being useful in applications that require single photons on demand at high repetition rates. However, the broad ZPL linewidths could pose a challenge for applications that require indistinguishable photons. Our work establishes the mechanisms responsible for linewidth broadening in these SPEs.

\section*{Methods}

The SPEs studied in this work are hosted in $\sim 4$ $\mu$m thick semi-insulating GaN epitaxial layers grown Ga-polar using HVPE on 430 $\mu$m thick sapphire substrates. The samples were obtained from PAM-XIAMEN Co.Ltd. A custom built confocal scanning microscope setup was used to optically excite the SPEs (using a 532 nm pump laser) and collect the PL. A $4f$ setup with a galvo mirror was used for scanning. The collected PL was split 50:50 into a spectrometer and a Hanbury-Brown and Twiss setup consisting of two photon counting detectors (PMA hybrid 40 from Picoquant) and a correlator (Multiharp150 from Picoquant). A 0.9 NA objective was used for all room temperature measurements, whereas for cryogenic temperature measurements, samples were mounted inside a cryostat and a 0.7 NA objective with a correction collar was used to collect PL through the cryostat window. The spectral resolution of the setup at $\sim$650 nm wavelength was $\sim$0.18 meV.

\bibliography{sample}

\begin{thebibliography}{10}
\urlstyle{rm}
\expandafter\ifx\csname url\endcsname\relax
  \def\url#1{\texttt{#1}}\fi
\expandafter\ifx\csname urlprefix\endcsname\relax\def\urlprefix{URL }\fi
\expandafter\ifx\csname doiprefix\endcsname\relax\def\doiprefix{DOI: }\fi
\providecommand{\bibinfo}[2]{#2}
\providecommand{\eprint}[2][]{\url{#2}}

\bibitem{aharonovich2016solid}
\bibinfo{author}{Aharonovich, I.}, \bibinfo{author}{Englund, D.} \&
  \bibinfo{author}{Toth, M.}
\newblock \bibinfo{journal}{\bibinfo{title}{Solid-state single-photon
  emitters}}.
\newblock {\emph{\JournalTitle{Nature Photonics}}}
  \textbf{\bibinfo{volume}{10}}, \bibinfo{pages}{631--641}
  (\bibinfo{year}{2016}).

\bibitem{claudon2010highly}
\bibinfo{author}{Claudon, J.} \emph{et~al.}
\newblock \bibinfo{journal}{\bibinfo{title}{A highly efficient single-photon
  source based on a quantum dot in a photonic nanowire}}.
\newblock {\emph{\JournalTitle{Nature Photonics}}}
  \textbf{\bibinfo{volume}{4}}, \bibinfo{pages}{174--177}
  (\bibinfo{year}{2010}).

\bibitem{santori2001triggered}
\bibinfo{author}{Santori, C.}, \bibinfo{author}{Pelton, M.},
  \bibinfo{author}{Solomon, G.}, \bibinfo{author}{Dale, Y.} \&
  \bibinfo{author}{Yamamoto, Y.}
\newblock \bibinfo{journal}{\bibinfo{title}{Triggered single photons from a
  quantum dot}}.
\newblock {\emph{\JournalTitle{Physical Review Letters}}}
  \textbf{\bibinfo{volume}{86}}, \bibinfo{pages}{1502} (\bibinfo{year}{2001}).

\bibitem{kurtsiefer2000stable}
\bibinfo{author}{Kurtsiefer, C.}, \bibinfo{author}{Mayer, S.},
  \bibinfo{author}{Zarda, P.} \& \bibinfo{author}{Weinfurter, H.}
\newblock \bibinfo{journal}{\bibinfo{title}{Stable solid-state source of single
  photons}}.
\newblock {\emph{\JournalTitle{Physical review letters}}}
  \textbf{\bibinfo{volume}{85}}, \bibinfo{pages}{290} (\bibinfo{year}{2000}).

\bibitem{neu2012photophysics}
\bibinfo{author}{Neu, E.}, \bibinfo{author}{Agio, M.} \&
  \bibinfo{author}{Becher, C.}
\newblock \bibinfo{journal}{\bibinfo{title}{Photophysics of single silicon
  vacancy centers in diamond: implications for single photon emission}}.
\newblock {\emph{\JournalTitle{Optics express}}} \textbf{\bibinfo{volume}{20}},
  \bibinfo{pages}{19956--19971} (\bibinfo{year}{2012}).

\bibitem{castelletto2014silicon}
\bibinfo{author}{Castelletto, S.} \emph{et~al.}
\newblock \bibinfo{journal}{\bibinfo{title}{A silicon carbide room-temperature
  single-photon source}}.
\newblock {\emph{\JournalTitle{Nature materials}}}
  \textbf{\bibinfo{volume}{13}}, \bibinfo{pages}{151--156}
  (\bibinfo{year}{2014}).

\bibitem{he2015single}
\bibinfo{author}{He, Y.-M.} \emph{et~al.}
\newblock \bibinfo{journal}{\bibinfo{title}{Single quantum emitters in
  monolayer semiconductors}}.
\newblock {\emph{\JournalTitle{Nature nanotechnology}}}
  \textbf{\bibinfo{volume}{10}}, \bibinfo{pages}{497--502}
  (\bibinfo{year}{2015}).

\bibitem{tran2016quantum}
\bibinfo{author}{Tran, T.~T.}, \bibinfo{author}{Bray, K.},
  \bibinfo{author}{Ford, M.~J.}, \bibinfo{author}{Toth, M.} \&
  \bibinfo{author}{Aharonovich, I.}
\newblock \bibinfo{journal}{\bibinfo{title}{Quantum emission from hexagonal
  boron nitride monolayers}}.
\newblock {\emph{\JournalTitle{Nature nanotechnology}}}
  \textbf{\bibinfo{volume}{11}}, \bibinfo{pages}{37--41}
  (\bibinfo{year}{2016}).

\bibitem{xue2020single}
\bibinfo{author}{Xue, Y.} \emph{et~al.}
\newblock \bibinfo{journal}{\bibinfo{title}{Single-photon emission from point
  defects in aluminum nitride films}}.
\newblock {\emph{\JournalTitle{The Journal of Physical Chemistry Letters}}}
  \textbf{\bibinfo{volume}{11}}, \bibinfo{pages}{2689--2694}
  (\bibinfo{year}{2020}).

\bibitem{berhane2017bright}
\bibinfo{author}{Berhane, A.~M.} \emph{et~al.}
\newblock \bibinfo{journal}{\bibinfo{title}{Bright room-temperature
  single-photon emission from defects in gallium nitride}}.
\newblock {\emph{\JournalTitle{Advanced Materials}}}
  \textbf{\bibinfo{volume}{29}}, \bibinfo{pages}{1605092}
  (\bibinfo{year}{2017}).

\bibitem{berhane2018photophysics}
\bibinfo{author}{Berhane, A.~M.} \emph{et~al.}
\newblock \bibinfo{journal}{\bibinfo{title}{Photophysics of gan single-photon
  emitters in the visible spectral range}}.
\newblock {\emph{\JournalTitle{Physical Review B}}}
  \textbf{\bibinfo{volume}{97}}, \bibinfo{pages}{165202}
  (\bibinfo{year}{2018}).

\bibitem{Nguyen2019point}
\bibinfo{author}{Nguyen, M.} \emph{et~al.}
\newblock \bibinfo{journal}{\bibinfo{title}{Effects of microstructure and
  growth conditions on quantum emitters in gallium nitride}}.
\newblock {\emph{\JournalTitle{APL Materials}}} \textbf{\bibinfo{volume}{7}},
  \bibinfo{pages}{081106} (\bibinfo{year}{2019}).

\bibitem{Nguyen2021polarity}
\bibinfo{author}{Nguyen, M. A.~P.} \emph{et~al.}
\newblock \bibinfo{journal}{\bibinfo{title}{Site control of quantum emitters in
  gallium nitride by polarity}}.
\newblock {\emph{\JournalTitle{Applied Physics Letters}}}
  \textbf{\bibinfo{volume}{118}}, \bibinfo{pages}{021103}
  (\bibinfo{year}{2021}).

\bibitem{sontheimer2017photodynamics}
\bibinfo{author}{Sontheimer, B.} \emph{et~al.}
\newblock \bibinfo{journal}{\bibinfo{title}{Photodynamics of quantum emitters
  in hexagonal boron nitride revealed by low-temperature spectroscopy}}.
\newblock {\emph{\JournalTitle{Physical Review B}}}
  \textbf{\bibinfo{volume}{96}}, \bibinfo{pages}{121202}
  (\bibinfo{year}{2017}).

\bibitem{lienhard2016bright}
\bibinfo{author}{Lienhard, B.} \emph{et~al.}
\newblock \bibinfo{journal}{\bibinfo{title}{Bright and photostable
  single-photon emitter in silicon carbide}}.
\newblock {\emph{\JournalTitle{Optica}}} \textbf{\bibinfo{volume}{3}},
  \bibinfo{pages}{768--774} (\bibinfo{year}{2016}).

\bibitem{Reineker1999T3}
\bibinfo{author}{Hizhnyakov, V.} \& \bibinfo{author}{Reineker, P.}
\newblock \bibinfo{journal}{\bibinfo{title}{Optical dephasing in defect-rich
  crystals}}.
\newblock {\emph{\JournalTitle{The Journal of chemical physics}}}
  \textbf{\bibinfo{volume}{111}}, \bibinfo{pages}{8131--8135}
  (\bibinfo{year}{1999}).

\bibitem{Fu2009JT}
\bibinfo{author}{Fu, K.-M.~C.} \emph{et~al.}
\newblock \bibinfo{journal}{\bibinfo{title}{Observation of the dynamic
  jahn-teller effect in the excited states of nitrogen-vacancy centers in
  diamond}}.
\newblock {\emph{\JournalTitle{Physical Review Letters}}}
  \textbf{\bibinfo{volume}{103}}, \bibinfo{pages}{256404}
  (\bibinfo{year}{2009}).

\bibitem{abtew2011dynamic}
\bibinfo{author}{Abtew, T.~A.} \emph{et~al.}
\newblock \bibinfo{journal}{\bibinfo{title}{Dynamic jahn-teller effect in the
  nv- center in diamond}}.
\newblock {\emph{\JournalTitle{Physical review letters}}}
  \textbf{\bibinfo{volume}{107}}, \bibinfo{pages}{146403}
  (\bibinfo{year}{2011}).

\bibitem{hizhnyakov2002zero}
\bibinfo{author}{Hizhnyakov, V.}, \bibinfo{author}{Kaasik, H.} \&
  \bibinfo{author}{Sildos, I.}
\newblock \bibinfo{journal}{\bibinfo{title}{Zero-phonon lines: the effect of a
  strong softening of elastic springs in the excited state}}.
\newblock {\emph{\JournalTitle{physica status solidi (b)}}}
  \textbf{\bibinfo{volume}{234}}, \bibinfo{pages}{644--653}
  (\bibinfo{year}{2002}).

\bibitem{Silsbee1962T7}
\bibinfo{author}{Silsbee, R.}
\newblock \bibinfo{journal}{\bibinfo{title}{Thermal broadening of the
  m{\"o}ssbauer line and of narrow-line electronic spectra in solids}}.
\newblock {\emph{\JournalTitle{Physical Review}}}
  \textbf{\bibinfo{volume}{128}}, \bibinfo{pages}{1726} (\bibinfo{year}{1962}).

\bibitem{jungwirth2016temperature}
\bibinfo{author}{Jungwirth, N.~R.} \emph{et~al.}
\newblock \bibinfo{journal}{\bibinfo{title}{Temperature dependence of
  wavelength selectable zero-phonon emission from single defects in hexagonal
  boron nitride}}.
\newblock {\emph{\JournalTitle{Nano letters}}} \textbf{\bibinfo{volume}{16}},
  \bibinfo{pages}{6052--6057} (\bibinfo{year}{2016}).

\bibitem{akbari2021temperature}
\bibinfo{author}{Akbari, H.}, \bibinfo{author}{Lin, W.-H.},
  \bibinfo{author}{Vest, B.}, \bibinfo{author}{Jha, P.~K.} \&
  \bibinfo{author}{Atwater, H.~A.}
\newblock \bibinfo{journal}{\bibinfo{title}{Temperature-dependent spectral
  emission of hexagonal boron nitride quantum emitters on conductive and
  dielectric substrates}}.
\newblock {\emph{\JournalTitle{Physical Review Applied}}}
  \textbf{\bibinfo{volume}{15}}, \bibinfo{pages}{014036}
  (\bibinfo{year}{2021}).

\bibitem{marseglia2011nanofabricated}
\bibinfo{author}{Marseglia, L.} \emph{et~al.}
\newblock \bibinfo{journal}{\bibinfo{title}{Nanofabricated solid immersion
  lenses registered to single emitters in diamond}}.
\newblock {\emph{\JournalTitle{Applied Physics Letters}}}
  \textbf{\bibinfo{volume}{98}}, \bibinfo{pages}{133107}
  (\bibinfo{year}{2011}).

\bibitem{jamali2014microscopic}
\bibinfo{author}{Jamali, M.} \emph{et~al.}
\newblock \bibinfo{journal}{\bibinfo{title}{Microscopic diamond
  solid-immersion-lenses fabricated around single defect centers by focused ion
  beam milling}}.
\newblock {\emph{\JournalTitle{Review of Scientific Instruments}}}
  \textbf{\bibinfo{volume}{85}}, \bibinfo{pages}{123703}
  (\bibinfo{year}{2014}).

\bibitem{neu2013low}
\bibinfo{author}{Neu, E.} \emph{et~al.}
\newblock \bibinfo{journal}{\bibinfo{title}{Low-temperature investigations of
  single silicon vacancy colour centres in diamond}}.
\newblock {\emph{\JournalTitle{New Journal of Physics}}}
  \textbf{\bibinfo{volume}{15}}, \bibinfo{pages}{043005}
  (\bibinfo{year}{2013}).

\bibitem{Santis2021stark}
\bibinfo{author}{De~Santis, L.}, \bibinfo{author}{Trusheim, M.~E.},
  \bibinfo{author}{Chen, K.~C.} \& \bibinfo{author}{Englund, D.~R.}
\newblock \bibinfo{journal}{\bibinfo{title}{Investigation of the stark effect
  on a centrosymmetric quantum emitter in diamond}}.
\newblock {\emph{\JournalTitle{Physical Review Letters}}}
  \textbf{\bibinfo{volume}{127}}, \bibinfo{pages}{147402}
  (\bibinfo{year}{2021}).

\bibitem{Cardona2001phonon}
\bibinfo{author}{Ruf, T.} \emph{et~al.}
\newblock \bibinfo{journal}{\bibinfo{title}{Phonon dispersion curves in
  wurtzite-structure gan determined by inelastic x-ray scattering}}.
\newblock {\emph{\JournalTitle{Physical review letters}}}
  \textbf{\bibinfo{volume}{86}}, \bibinfo{pages}{906} (\bibinfo{year}{2001}).

\bibitem{Ziman}
\bibinfo{author}{Ziman, J.~M.}
\newblock \emph{\bibinfo{title}{Electrons and Phonons}}
  (\bibinfo{publisher}{Oxford University Press}, \bibinfo{year}{1960}),
  \bibinfo{edition}{1st} edn.

\bibitem{Milburn}
\bibinfo{author}{Walls, D.} \& \bibinfo{author}{Milburn, G.~J.}
\newblock \emph{\bibinfo{title}{Quantum Optics}} (\bibinfo{publisher}{Springer
  Verlag}, \bibinfo{year}{2008}), \bibinfo{edition}{2nd} edn.

\bibitem{Phillips2011}
\bibinfo{author}{Callsen, G.} \emph{et~al.}
\newblock \bibinfo{journal}{\bibinfo{title}{Phonon deformation potentials in
  wurtzite gan and zno determined by uniaxial pressure dependent raman
  measurements}}.
\newblock {\emph{\JournalTitle{Applied Physics Letters}}}
  \textbf{\bibinfo{volume}{98}}, \bibinfo{pages}{061906}
  (\bibinfo{year}{2011}).

\end{thebibliography}

\section*{Acknowledgements}

This work was supported by the Cornell Center for Materials Research with funding from the NSF MRSEC program (DMR-1719875) and also by the NSF-RAISE:TAQS (ECCS-1838976).

\section*{Appendix: Fitting a $T^{3}$ power law to the ZPL linewidth data}
\begin{figure}[tb]
        \centering
		\includegraphics[width=0.4\linewidth]{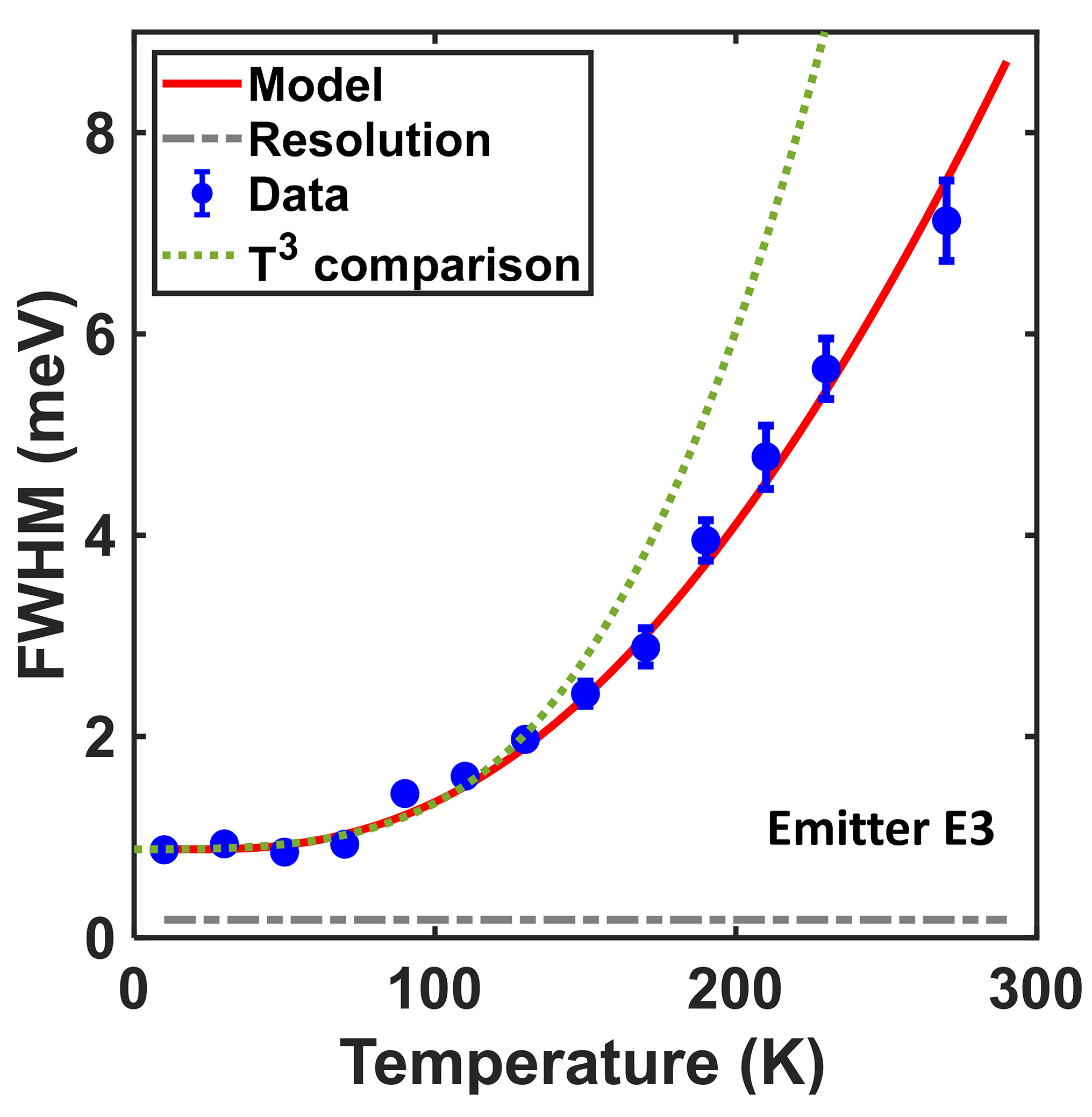}
		\caption{The FWHM linewidth of emitter E3 is plotted as a function of the temperature. The solid line is the fit to the data using the theoretical model discussed in the text. The dotted line shows an attempt to fit the data with a model in which the temperature dependence of the linewidth is proportional to $T^{3}$.}
		\label{fig:app}
\end{figure}

Fig.~\ref{fig:app} shows an attempt to fit the ZPL FWHM linewidth data of emitter E3 with a model in which the temperature dependence of the linewidth $f_{L}$ of the Lorentzian, determined by the decoherence rate $\gamma$, is $C T^{3}$, where $C$ is a constant. If the value of $C$ is adjusted to fit the data well at low temperatures then the $CT^{3}$ model predicts an increase in the linewidth with the temperature that is much steeper than the data, as shown in Fig.~\ref{fig:app}. It is obvious from this that decoherence mechanisms that give $T^{5}$ or $T^{7}$ temperature dependence will also not agree with our data.  

\end{document}